\documentclass[]{article}
\usepackage[applemac]{inputenc}	
\usepackage[english]{babel}
\usepackage{graphicx}
\usepackage{epsfig}
\usepackage{epstopdf}
\usepackage{psfrag}
\usepackage{subfig}
\usepackage{latexsym}
\usepackage{amsmath}
\usepackage{bm}
\usepackage{amssymb}
\usepackage{siunitx}
\usepackage{upgreek}
\usepackage{color}
\usepackage{xcolor}
\usepackage{comment}
\usepackage{natbib}

\newcommand{\reff}[2][]{\ref{fig:#2}{\it #1}} 
\newcommand{\refe}[1]{\eqref{eq:#1}} 
\newcommand{\refs}[1]{\ref{sec:#1}} 

\usepackage{xspace}
\newcommand{\nsd}[1]{\textit{#1}\xspace} 

\newcommand{\re}{\nsd{Re}}

\newcommand{\abs}[1]{\lvert#1\rvert}

\newcommand{\mum}{$\upmu{}$m\xspace}
\newcommand{\mus}{$\upmu{}$s\xspace}

\newcommand{\phic}{\phi_c}
\newcommand{\Deq}{D_{eq}}

\usepackage{authblk}
\title{Deformation upon impact of a \break concentrated suspension drop}
\author[]{Loren J\o{}rgensen, Yo\"{e}l Forterre \& Henri Lhuissier\thanks{Corresponding author: \texttt{henri.lhuissier@univ-amu.fr}}}
\affil{Aix Marseille Universit\'e, CNRS, IUSTI, Marseille, France\\}
\date{\today}

\begin{document}
\maketitle

\begin{abstract}
	We study the impact between a plate and a drop of non-colloidal solid particles suspended in a Newtonian liquid, with a specific attention to the case when the particle volume fraction, $\phi$, is close to — or even exceeds — the critical volume fraction, $\phic$, at which the steady effective viscosity of the suspension diverges. We use a specific concentration protocol together with an accurate determination of $\phi$ for each drop and we measure the deformation $\beta$ for different liquid viscosities, impact velocities and particle sizes. At low volume fractions, $\beta$ is found to follow closely an effective Newtonian behavior, which we determine by documenting the low deformation limit for a highly viscous Newtonian drop and characterizing the effective shear viscosity of our suspensions. By contrast, whereas the effective Newtonian approach predicts that $\beta$ vanishes at $\phic$, a finite deformation is observed for $\phi>\phic$. This finite deformation remains controlled by the suspending liquid viscosity and increases with increasing particle size, which suggests that the dilatancy of the particle phase is a key factor of the dissipation process close to and above $\phic$.
\end{abstract}

\section{Introduction}\label{sec:intro}

	The collision between a particle-laden drop and a solid is encountered in many natural situations or industrial applications, such as raindrop impacts on granular soils \citep{Zhao2015}, spores dispersal \citep{Ingold1971} or ink-jet printing of colloidal particles drop for microelectronics \citep{Korkut2008,Qi2011} and microfabrication \citep{Derby2011}. It is however a challenging problem, which couples the dynamics of a highly unsteady free surface flow with the complex rheology of particulate suspensions. 
	For a Newtonian liquid the impact dynamics may involve capillarity or the influence of ambient air, but for a sufficiently viscous medium the deformation of the drop is only controlled by a balance between the drop initial kinetic energy and bulk dissipation \citep{Madejski76,Chandra91,Laan14,Josserand16,Wildeman16,Gordillo18}. This over-damped limit may seem relevant for the impact of a drop containing a large quantity of solid particles, since the bulk effective viscosity of a particulate suspension increases strongly with increasing volume fractions and eventually diverges at the critical volume fraction $\phic\approx 0.60$ \citep{Zarraga00,Stickel05,Guazzelli18}. However, depending on the preparation protocol, the particle volume fraction of a suspension can actually exceed $\phic$. In such case the effective viscosity is undefined and it remains unclear how much an impacting drop spreads for $\phi>\phic$, as well as whether a Newtonian effective approach is relevant when $\phi$ becomes close to $\phic$ or what is the influence of the particle size.
\begin{figure}\begin{center}
	\includegraphics[width=1\linewidth]{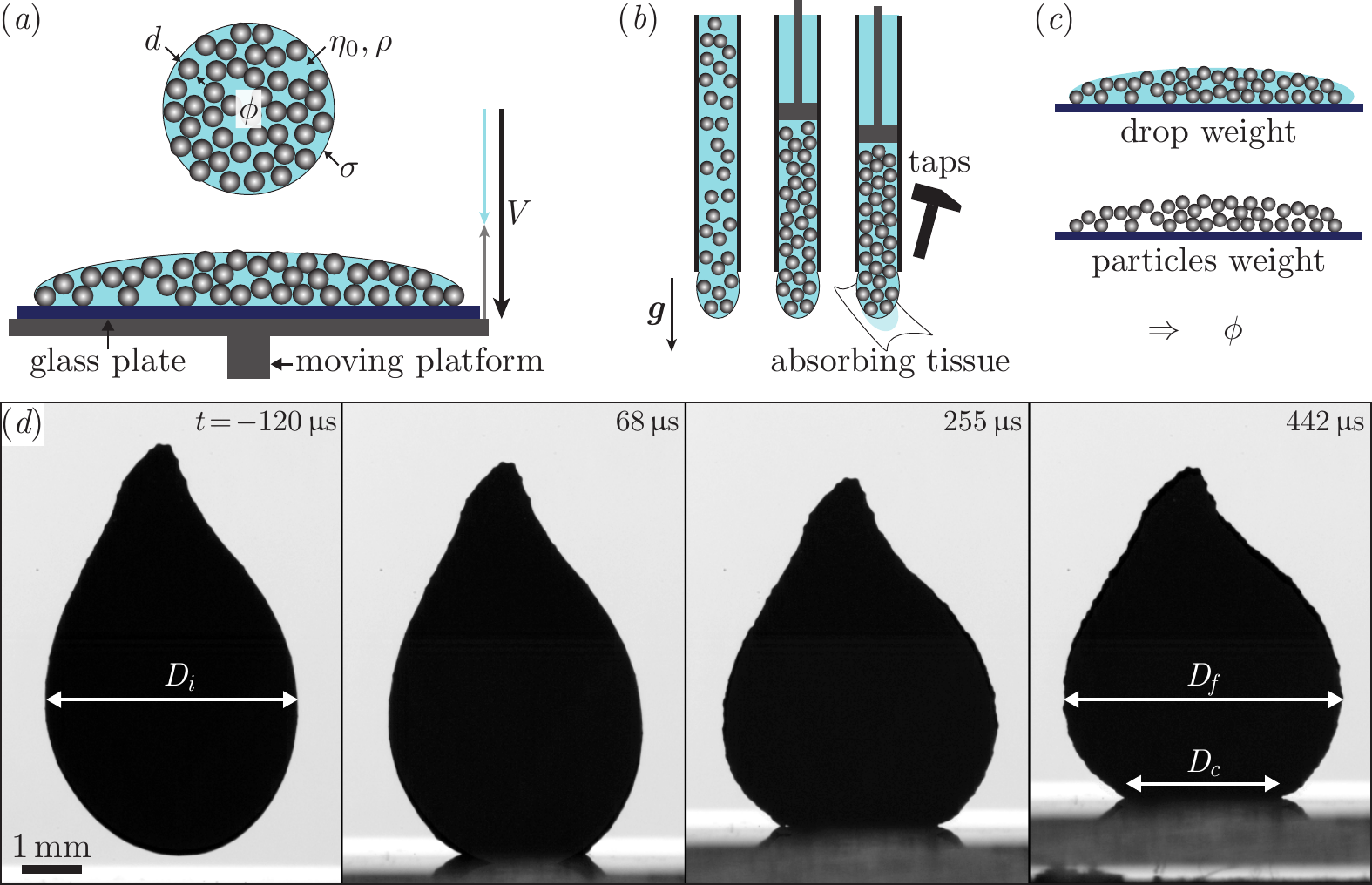}
	\caption{({\it a-c}) Schematics of ({\it a}) the impact setup, ({\it b}) the concentrated drop production method and ({\it c}) the post-impact measurement of the particle volume fraction. ({\it d}) Spreading dynamics for a drop with a particle volume fraction $\phi = 0.556$ and an impact velocity $V= 4$\,m/s ($d=145$\,\mum, $\eta_0 = 50.5\,$mPa\,s, $t=0$ indicates the contact instant).}
	\label{fig:setup}
\end{center}\end{figure}

	So far, the limit of high volume fractions have been studied mostly for suspensions of small particles subjected to colloidal forces, which eventually shear-thicken 
\citep{Bertola15,Boyer16}, or to document the capillary limited spreading and splashing for non-neutrally buoyant particles \citep{Peters13, Lubbers14, Schaarsberg16, Raux20}. To our knowledge, the viscous spreading of drops of non-colloidal particles has only been studied up to intermediate volume fractions ($\phi< 0.45$) with a Newtonian effective approach assuming a large deformation of the drop \citep{Nicolas05}. The viscous spreading of a highly concentrated suspension drop has thus remained an open question, which is precisely the concern of this paper.
We tackle the question by focusing on high-velocity impacts (large Froude and Weber numbers) and over-damped spreadings and studying the early dynamics during which the initial kinetic energy of the drop is dissipated (i.e., before any slower gravitational/capillary relaxation occurs). In \S\refs{exp}, we present the experimental protocol used to prepare and impact highly concentrated drops, as well as first observations for various volume fractions, impact velocities and viscosities. In \S\refs{Newt}, we analyse these results in the framework of a viscous effective modeling of the suspension, which is calibrated with experiments on Newtonian drops. Finally, some crucial limitations of the effective viscosity approach are evidenced for particle volume fractions close to or above $\phi_c$, which are discussed in terms of granular dilatancy effects and their coupling with the liquid pore pressure (\S\refs{limit}).


\section{Experimental setup and general observations}\label{sec:exp}

\subsection{Setup}\label{sec:setup}

	The drop impact protocol is sketched in figure \reff{setup}. It consists in letting a suspension drop fall and impact perpendicularly onto a plate and observing the first milliseconds following the impact, during which the kinetic energy is dissipated.
		
	The suspensions are prepared by immersing spherical polystyrene particles (Dynoseed TS, Microbeads, with a density of 1049\,kg/m$^{3}$) in a Newtonian liquid with a close density. Most of the experiments are performed with a mean particle diameter $d=145\,$\mum and a relative standard deviation in size below 4$\%$. A larger diameter ($d=585\,$\mum$ \pm 1\%$) is also used (see \S\refs{limit}). The liquid is a water solution of PEGPG (3.9\,kg/mol poly(ethylene glycol-ran-propylene glycol)-monobutylether by Sigma-Aldricht), which wets the particles well and is sufficiently viscous to avoid that particles are ejected from the drop during the impact. Its viscosity, $\eta_0$, is varied between 0.93 and 192\,mPa\,s by adjusting the PEGPG concentration (from 0 to 50wt$\%$), whereas its density and surface tension vary from 998 to 1056\,kg/m$^3$ and from 70 to 40\,mN/m, respectively.

	The particle volume fraction of the drop, $\phi$, is varied from 0 to above the critical volume fraction. For moderate concentrations ($\phi\leq0.54$) the formation of the drop is straightforward. The suspension is prepared at the desired $\phi$, infused in a truncated (nozzle-free) plastic syringe with an inner diameter of $4.6\,$mm and the suspension is slowly extruded until the drop detaches. For $\phi<0.54$, this simple protocol ensures that the volume fraction in the drop is identical (within less than 0.01, as verified a posteriori) to the nominal volume fraction of the preparation. However, it does not allow to control and reproduce the solid content for more concentrated drops. Those drops, with $\phi>0.54$, are thus obtained by infusing a suspension with a volume fraction of $\approx 0.55$ in the syringe and concentrating each drop inside the syringe, by letting the particles sediment under gentle tapping and soaking part of the liquid with an absorbing tissue (see figure \reff{setup}{\it b}). In this case the value of $\phi$ is determined post impact by weighing, for each single drop, both the whole drop and its dry content obtained after washing and drying the particles. This protocol permits to reach volume fractions as high as 0.62 with an uncertainty of less than 0.005. The most concentrated drops (above $\phic \simeq 0.605$) behave like a paste. To avoid that they extrude as a long cylinder their shape is progressively relaxed by gently tapping the syringe. The detached drops remain oblong (see figure \reff{setup}{\it d}) and care is taken to discard the impacts for which the drop has rotated. 
	
	The drop impacts a smooth glass plate (washed and rinsed with micro-filtrated water before each impact). However, a series of impacts has also been performed on a rough surface (flat sand paper with a grain size of $\approx 160\,$\mum $\approx d$) to verify the influence of a potential sliding of the particles on the surface (see \S\refs{obs}). The impact velocity $V$ is varied from 1.6 to 10\,m/s. In order to achieve the largest velocities while keeping aerodynamic effects on the drop negligible the drop is released from a small height ($\approx 20$\,cm) and the plate is moved vertically towards the drop by a fast translating stage. The stage motion is triggered by the occulting of a laser beam by the falling drop and is tuned so as to insure that the impact occurs at the fixed observation spot with the desired impact velocity $V$ ($=\abs{V_{\rm drop}}+\abs{V_{\rm stage}}$). The impact is imaged with a fast camera facing an intense collimated light source, which yields a sufficiently short time-sampling (62\,\mus with a shutter time of 2\,\mus) and high spatial resolution (16\,\mum). A grazing view of the plate is used in order to access the whole deformation of the drop, including at the contact with the plate.

\subsection{General observations}\label{sec:obs}

	A typical impact dynamics for a concentrated suspension drop is illustrated in figure \reff{setup}{\it d} for $\phi = 0.556$, $\eta_0=50.5\,$mPa\,s and $V=4$\,m/s. As the drop makes contact with the plate its lower surface flattens and the remaining of the drop widens while contracting in the vertical direction. The deformation is initially restricted to the neighbouring of the plate and progressively extends up to the top of the drop. The whole sequence lasts only a few hundreds of microseconds, after which all the kinetic energy is dissipated and fast deformations stop (the subsequent capillary-gravitational relaxation of the shape, lasting $\gtrsim 1$\,s, is orders of magnitude slower).
\begin{figure}\begin{center}
	\includegraphics[width=1\linewidth]{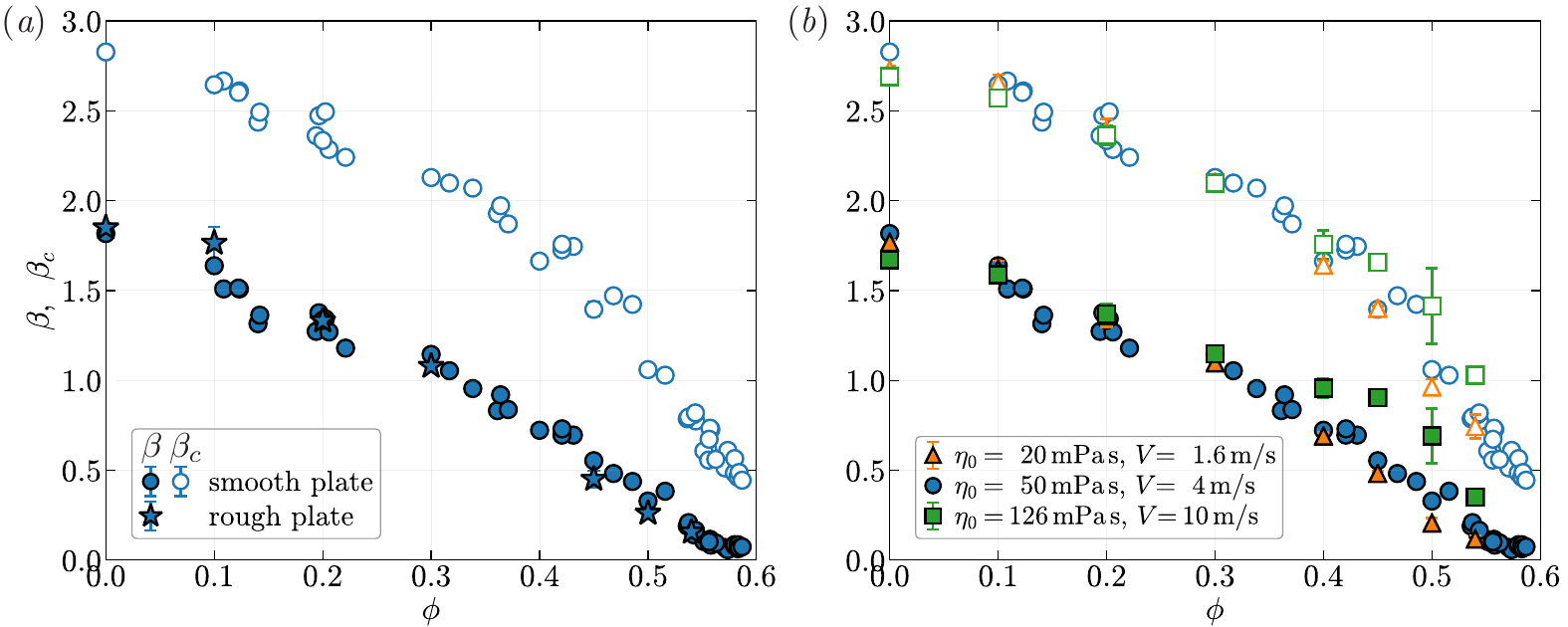}
	\caption{({\it a}) Relative spreading as a function of the particle volume fraction for a suspension drop with $d=145\,\upmu$m and $\eta_0 = 50.5\,$mPa\,s ($V= 4$\,m/s), as measured from the maximal deformed diameter $D_f$ ($\beta\equiv D_f/D_i-1$) and from the contact diameter $D_c$ ($\beta_c\equiv D_c/D_{eq}$). ({\it b}) Spreading versus $\phi$ for different suspending liquids and the same ratio $V/\eta_0$.}
	\label{fig:obs}
	\vspace*{-5mm}
\end{center}\end{figure}
	
	We characterize the final spreading by measuring both the maximal diameter $D_f$ and the contact diameter $D_c$ (as observed from the side view, see figure \reff{setup}{\it d}). This yields two different measurements of the deformation:
\begin{eqnarray}
	 \beta \equiv \frac{D_f}{D_i}-1, \quad\quad\text{and}\quad\quad \beta_c \equiv \frac{D_c}{D_{eq}},
	 \label{eq:beta}
\end{eqnarray}
where $D_i$ and $D_{eq}$ are, respectively, the maximal diameter before impact and the diameter of the equivalent volume sphere. These definitions satisfy $\beta=\beta_c=0$ for a vanishing deformation. To obtain the fairest comparison between the slightly different drop shapes, as well as with the spherical geometry assumed below, we choose to scale $D_c$ by $D_{eq}$ rather than $D_i$ (which cannot be done for $\beta$ since we require that it vanishes in the absence of deformation).

	The dependance of the spreading on the particle volume fraction is presented in figure \reff{obs}{\it a}, for the same suspension and impact velocity as in figure \reff{setup}{\it d}. For both $\beta$ and $\beta_c$ the spreading decreases continuously and significantly from $\phi =0$ up to the largest concentration explored here ($\phi = 0.587$). Importantly, the spreading is found to be almost unchanged upon using an impact plate with roughnesses of the order of $d$ rather than a smooth glass plate. This suggests that, even with the smooth plate, there is no significant slipping of the particles at the plate surface.
	
	Before seeking to understand the dependence to the particle content, a last crucial remark can be drawn by comparing the deformation obtained by varying both the velocity and suspending liquid such that the ratio $V/\eta_0$ remains constant, i.e., keeping the Reynolds number at fixed volume fraction unchanged. As shown in figure \reff{obs}{\it b}, the deformation is found to be essentially unchanged for a 6-fold increase in $V$ (with $\re_0\equiv \rho V\! D_i /\eta_0=350\pm20$) over the whole range of $\phi$. This suggests that, similarly to the Newtonian case and in contrast with the Bagnold scaling invoked in \cite{Boyer16}, the dissipation of the motion of the suspension drops follows a viscous scaling and is directly controlled by the Reynolds number.
			
	To go beyond these findings and precisely assess the similarity and differences with the Newtonian case, a clear Newtonian reference is required, in particular in the limit of large viscosities (low Reynolds numbers), where the deformation is expected to be small.


%
%

\section{Newtonian effective approach}\label{sec:Newt}

\subsection{Newtonian calibration and low deformation limit}\label{sec:calib}

\begin{figure}\begin{center}
	\includegraphics[width=1\linewidth]{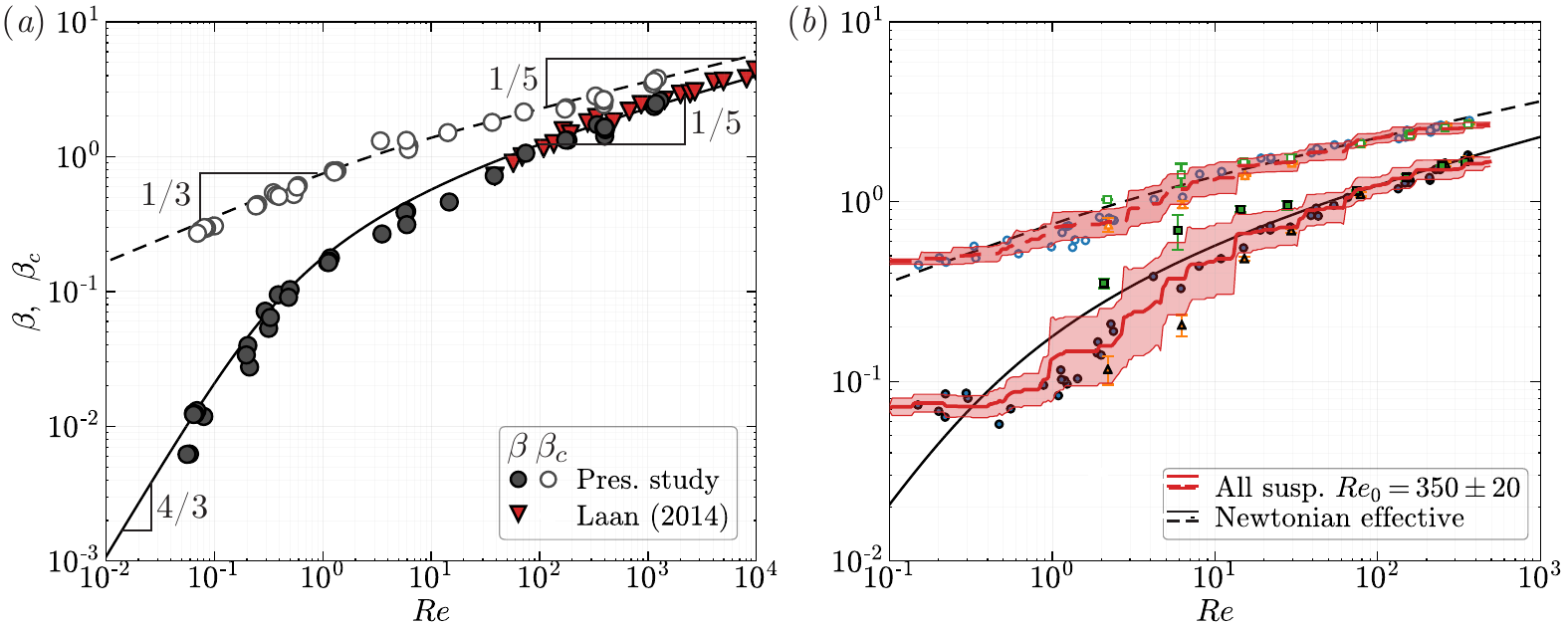}
	\caption{({\it a}) Spreading as a function of the impact Reynolds number for a viscous Newtonian drop. The red triangles represent the measurements of \cite{Laan14}. The lines represent equation \refe{Ffit}. ({\it b}) Same data as in figure \reff{obs}{\it b} plotted versus the effective Reynolds number $\rho D_iV/\eta(\phi)$ (resp. $\rho \Deq V/\eta(\phi)$) for $\beta$ (resp. $\beta_c$) defined according to equation \refe{eta} with the critical volume fraction, $\phi_c = 0.605$, obtained with the indentation setup (figure\,\reff{rheo}). The red line (resp. envelop) represents a sliding geometrical average (resp. standard deviation) of the spreading for the suspensions.} 
	\label{fig:Newt}
	\vspace*{-5mm}
\end{center}\end{figure}
	The Newtonian reference is obtained by impacting drops of Newtonian liquids (PEGPG and Ucon Oil aqueous solutions, with a viscosity $\eta$ between 0.012 and 41.5\,Pa\,s) for values of $\re = \rho D_i\!V/\eta$ ranging from 10$^{-1}$ to 10$^{3}$. As shown in figure \reff{Newt}{\it a}, for large $\re$ we recover the large deformation limit, $\beta \sim \re^{1/5}$, derived by \cite{Madejski76}. However, for larger viscosities the deformation is so small that it is restricted to the surrounding of the contact with the plate. In this case we expect a different law. Assuming, in the spirit of Hertz's model for elastic deformations \citep{Hertz81}, that the deformed drop remains close to a sphere except for the flattened contact area with diameter $D_c\ll D_f \simeq D_i$ where it is indented with a typical depth $\delta\sim D_c^2/D_i$, one expects that the deformation is restricted to a volume $\Omega\sim D_c^3$ and has a typical magnitude $\varepsilon\sim \delta/D_c$. Equating the initial kinetic energy of the drop, $\sim \rho V^2 D_i^3$, with the viscous dissipation, $\sim\eta_0\dot{\varepsilon} \varepsilon\Omega\sim \eta_0\varepsilon^2\Omega/(\delta/V)$, gives us $D_c/D_i\sim \re^{1/3}$. Last, making use of mass conservation, which imposes that the volume increase of the spherical part of the drop, ${D_f}^3-{D_i}^3\sim {D_i}^2(D_f-D_i)$, be equal to the indented volume, $\sim\delta D_c^2$, we obtain
\begin{eqnarray}
	\beta \equiv \frac{D_f}{D_i}-1 \sim \re^{4/3} \quad\text{and}\quad   \beta_c \equiv\frac{D_c}{D_{eq}} \sim \re^{1/3} \quad\text{for}\quad \re\ll1,
	\label{eq:F}
\end{eqnarray}
which is found to be in good agreement with the measurements for small deformations (figure \reff{Newt}{\it a}). Note that, although the most viscous drops do not relax to a spherical shape before impact (similarly to the concentrated suspension drops), the two different definitions of the deformation yield the same agreement between equation \refe{F} and the measurements, which suggests that the small departures to sphericity do not affect significantly the characterization of the deformation.

	In order to allow the comparison between the suspension case and the Newtonian case for any deformation we fit the Newtonian data with the following cross-over functions
\begin{eqnarray}
	\re &\equiv& F^{-1}(\beta) = \left[(2\beta)^{3/16}+(1.4\beta)^{5/4}\right]^4,\nonumber\\
	\quad\text{and}\quad \re &\equiv& {F_c}^{-1}(\beta_c) = \left[(1.3\beta_c)^6+(1.1\beta_c)^{10}\right]^{1/2},
	\label{eq:Ffit}
\end{eqnarray}
which are plotted in figure \reff{Newt}{\it a} together with the measurements. Their functional form is chosen so as to recover the scaling laws of both the large and the low deformation limits. 

	In order to serve as a reference for the suspension case, this Newtonian formalism has to be complemented with a characterization of the effective rheology of the suspensions.

\subsection{Rheological characterization of the suspensions}\label{sec:rheo}

\begin{figure}\begin{center}
	\includegraphics[width=1\linewidth]{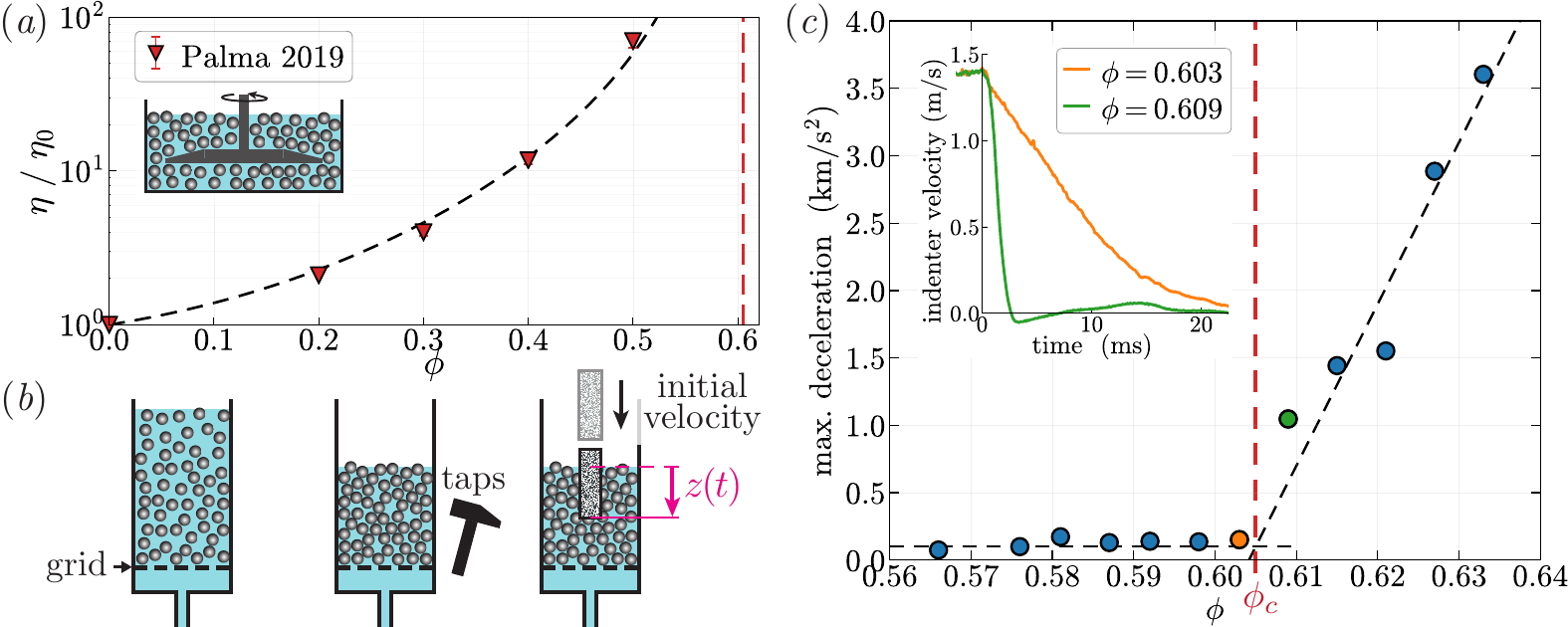}
	\caption{Rheological characterization of the suspensions. ({\it a}) Relative effective viscosity measured by \cite{Palma19} for the same suspensions ($d=145\,$\mum{}, PEGPG solution) as those used in the present study. The black dashed line is equation \refe{eta} with $a=2.2$. The red dashed line indicates $\phic$ as measured in ({\it b-c}). ({\it b}) Indentation setup used to measure the critical volume fraction $\phic$. ({\it c}) Inset: Decay of the indenter initial velocity for slightly different volume fractions of the sedimented particle pile  ($t=0$ indicates the contact instant). Main: maximal post-impact deceleration versus $\phi$. The horizontal and inclined dashed lines are fitted to the data. Their intersection sets $\phic\simeq0.605$.}
	\label{fig:rheo}
	\vspace*{-5mm}
\end{center}\end{figure}

	Figure \reff{rheo}{\it a} reports the bulk effective shear viscosity as a function of the particle volume fraction up to $\phi =0.50$, as measured in one of our previous works  for the same particles and liquids \citep[][the sketch of the dedicated shear cell in shown as an inset]{Palma19}. To characterize the suspension at much higher volume fractions that cannot be accessed with the shear cell, we use a complementary protocol inspired by \cite{Jerome16}: a protocol of dynamic indentation into a sedimented state of the suspension, which allows us to measure accurately the critical volume fraction $\phic$ by probing the transient rheology both below and above $\phic$. A sedimented particle pile is prepared at a volume fraction between $\simeq 0.565$ and $\simeq 0.632$ by letting the particle sediment gently at the loose random packing fraction $\simeq 0.565$ and increasing the packing by a controlled tapping on the container (see figure \reff{rheo}{\it b}). The pile is indented dynamically with a solid cylinder impacting at $1.4\,$m/s. The post-impact velocity of the indenter is measured (with high precision by applying conventional image correlation methods to the speckle-patterned surface of the indenter), as shown in the inset of figure \reff{rheo}{\it c}. For each value of $\phi$, the maximal deceleration of the indenter is extracted. As shown in figure \reff{rheo}{\it c}, the deceleration is almost constant (at typically 100\,m/s$^2$) up to $\phi \simeq 0.605$, and increases significantly and almost linearly (by a factor up to 40), above. This steep change in the transient rheology of the suspension indicates the onset of a dilatant response to deformation \citep{Jerome16}, which implies an enhanced dissipation because the increase in pore volume between the particles forces a Darcy flow of the suspending liquid through the particle phase. By definition, the onset of dilatancy is also the maximal volume fraction at which a constant-volume deformation is possible, i.e., the volume fraction $\phic$ at which the steady effective rheology of the suspension diverges. We therefore obtain a direct and accurate characterization of $\phic$ for our suspensions by determining the onset of the dilatancy from the fitting procedure defined in the caption of figure \reff{rheo}{\it c}, which yields $\phic\simeq0.605$. This exhaustive characterization is obtained for $d=145\,$\mum in pure water. However, by measuring the volume fraction of the loose random packing obtained by slow sedimentation (without tapping) we could verify that the value is almost the same (between 0.554 and 0.567) for all the different particles and suspending liquids that we have used, which indicates that the values of $\phic$ should also be almost the same. Similarly quasi-static measurements of the starting angle of avalanche in rotating drum yielded very similar values for particles in water (25.5°) and in a 25w$\%$ PEGPG solution (27°). 

	To obtain an effective viscosity reference for the whole range of $\phi$ we choose to summarize the above rheological characterizations by the following functional form
\begin{eqnarray}
	\frac{\eta(\phi)}{\eta_0} = 1+a-\frac{5\phic}{2}+\left(\frac{5\phic}{2}-2a\right)\left[1-\frac{\phi}{\phic}\right]^{-1}+a\left[1-\frac{\phi}{\phic}\right]^{-2},
	\label{eq:eta}
\end{eqnarray}
where $a$ is a fitting parameter. Similarly to Eilers' law \citep{Stickel05}, equation \refe{eta} verifies both Einstein relation ($\eta/\eta_0\rightarrow 1+5\phi/2$) in the diluted limit and the inverse quadratic divergence ($\eta/\eta_0\propto (\phic-\phi)^{-2}$) reported experimentally near $\phic$. This ensures that equation \refe{eta} (with the fitted value $a=2.2$) accurately represents the experimental data over the range $0\leq\phi\leq0.50$, as shown in figure \reff{rheo}{\it a}, but also close to the critical volume fraction $\phic$.

\subsection{Newtonian effective deformation}\label{sec:spread}

	From the viscous spreading law, $\beta(\rho V\! D_i/\eta)$ \refe{Ffit}, and the model for the steady effective viscosity of the suspensions, $\eta(\phi)$ \refe{eta}, we obtain a prediction for the Newtonian effective spreading of a suspension drop as a function of the particle volume fraction: 
\begin{eqnarray}
	\beta = F\left(\frac{\eta_0}{\eta(\phi)} \re_0\right) \qquad\text{and}\qquad  \beta_c = F_c\left(\frac{\eta_0}{\eta(\phi)} \re_0\right),
	\label{eq:betaN}
\end{eqnarray}
with $\re_0 = \rho V\! D_i/\eta_0$. From the monotonic increase of $\beta(\re)$ and $\eta(\phi)$ and their low Reynolds number limits ($\beta \sim \re^{4/3}$, $\beta_c \sim \re^{1/3}$ and $\eta \propto (\phic-\phi)^{-2}$), equation \refe{betaN} predicts that the deformation decreases with increasing $\phi$ and vanishes rapidly as the particle volume fraction approaches $\phic$, according to $\beta \propto (\phic-\phi)^{8/3}$ and $\beta_c \propto (\phic-\phi)^{2/3}$. Consistently, equation \refe{betaN} implies that the drop is not deformed at all above $\phic$ since the steady effective viscosity is infinite above $\phic$. 

	We are now able to assess the relevance of the effective Newtonian approach to the suspension case. Figure \reff{Newt}{\it b} compares the same data as in figure \reff{obs} to equation \refe{betaN}. At low $\phi$ (large deformations and effective Reynolds numbers) the spreading of the suspensions is found fo follow closely the Newtonian effective law. By contrast, a significant departure is observed on the low deformation side when the volume fraction approaches $\phic$. Instead of continuing to decrease with decreasing effective Reynolds number, the spreading of the suspension seems to plateau or decrease much more slowly (although the spreading of purely Newtonian drops is still experimentally measurable and in agreement with the Newtonian law over this range of deformations; see figure \reff{Newt}{\it a}).

\section{Dilatancy effects and influence of the particle size}\label{sec:limit}

\begin{figure}\begin{center}
	\includegraphics[width=1\linewidth]{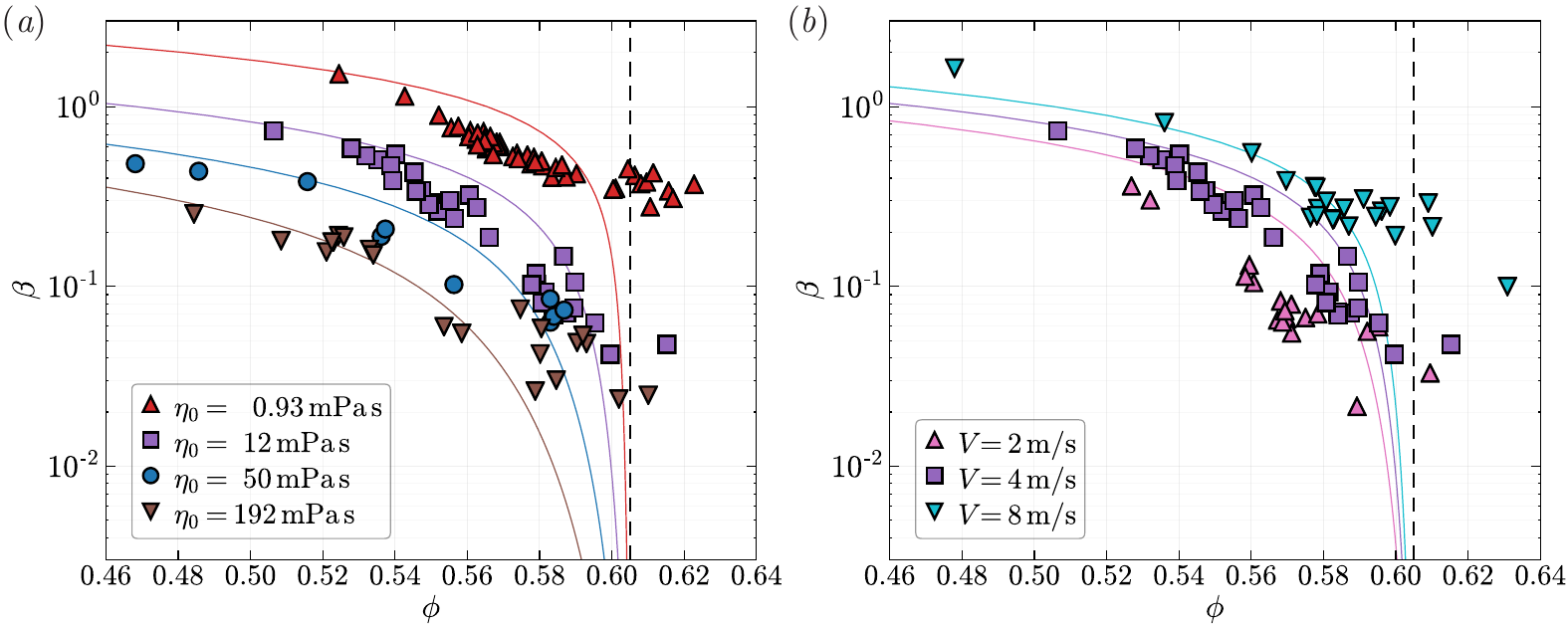}
	\caption{Spreading versus $\phi$ around the critical volume fraction $\phic$ for different viscosities of the suspending liquid ({\it a}, $V=4\,$m/s) and varied impact velocities ({\it b}, $\eta_0 = 12.0\,$mPa\,s). The particle size is $d = 145\,\upmu$m. The colored solid lines show the viscous effective law \refe{beta}. The vertical dashed line indicates $\phi_c$.}
	\label{fig:highphi}
	\vspace*{-5mm}
\end{center}\end{figure}
	The measurements discussed above indicate a deviation from the effective Newtonian behavior as $\phi$ becomes close to $\phi_c$. Yet, although the deformation is not predicted any more by an effective Newtonian approach, experiments with volume fractions increased beyond $\phic$ confirm that the deformation remains controlled by a balance between inertia and viscous dissipation. As shown in figures \reff{highphi}{\it a}-{\it b}, the spreading remains a decreasing function of the suspending liquid viscosity $\eta_0$ and an increasing function of the impact velocity $V$, including for $\phi$ above $\phic$, where a finite deformation is observed in stark contrast with the Newtonian effective expectation of a zero deformation.

	The above observations suggest that at least one important aspect of the impact that is neglected by the Newtonian effective approach has to be considered to explain the spreading of very concentrated drops. Upon the impact, the deformation of the particle phase, which determines the dissipation inside the suspension, does not necessarily proceed at a fixed volume fraction, in the sense of a fixed average distance between the particles. The volume of liquid and particles is conserved, but the particle phase can actually dilate by protruding from the liquid. As known since Reynolds' seminal demonstration of dilatancy \citep{Reynolds85}, this is actually the only way by which the drop can deform if it contains rigid particles at a volume fraction above $\phic$. Direct observation confirms that the interface of the suspension becomes qualitatively more corrugated as the drop touches the plate and deforms (see, e.g., figure \reff{setup}{\it d}), but we are not able to quantify the effect because the protrusions are masking each other in the images.
\begin{figure}\begin{center}
	\includegraphics[width=1\linewidth]{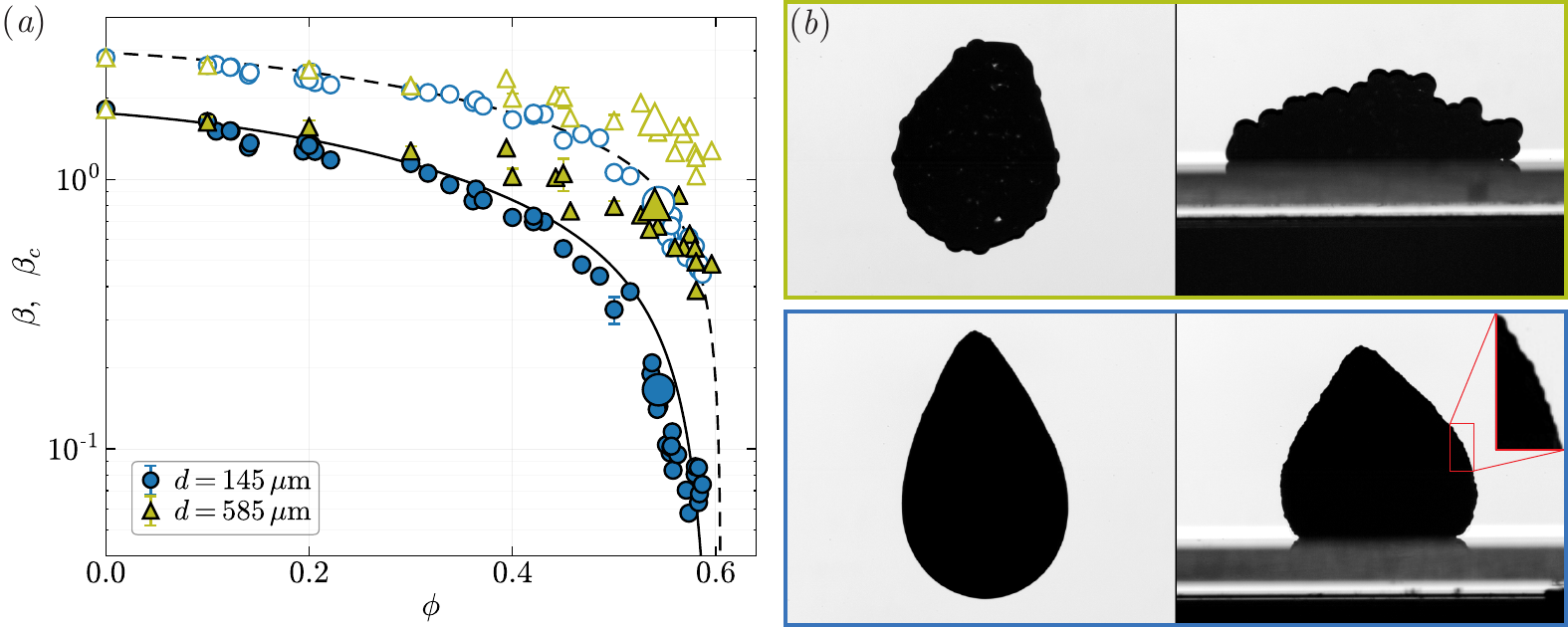}
	\caption{Influence of the particle size for the same suspending liquid and impact parameters ($\eta_0 = 50.5\,$mPa\,s, $V = 4\,$m/s). ({\it a}) Spreading versus $\phi$ (same data as in figure \reff{obs}{\it a} for $d=145\,$\mum). ({\it b}) Comparison of the drop deformations for the same volume fraction $\phi =0.54$ (larger symbols in ({\it a})).}
	\label{fig:d}
\end{center}\end{figure}
	
	Nevertheless, we can test the expected consequences of the dilatant motion. To fill the increasing volume between the diverging particles the latter requires a Darcy flow of the suspending liquid through the particle phase \citep{Jerome16}. For a dilatancy with a volume fraction decrease $-\Delta\phi\ll\phi$ one expects a dissipation per unit volume $\sim P\Delta\phi/\phi$, where the pore pressure $P$ built up by the Darcy flow follows $-P/D_i^2 \sim (1/\kappa)\eta_0 \dot{\varepsilon}(\Delta \phi/\phi)\sim (1/\kappa)\eta_0 V/D_i(\Delta \phi/\phi)$ and $\kappa\sim 10^{-3}d^2$ is the typical permeability of the particle phase at high solid fractions. In the limit of very small deformations and $\phi\gtrsim\phic$, the dilatancy may be assumed to be slaved to the deformation, i.e., $-\Delta \phi/\phi\sim\beta$, and the deformation should be limited by the dissipation of the dilatant motion, i.e., $P\Delta\phi/\phi \sim \rho V^2$. This assumption is found to be compatible with experiments. For the experimental values $d=145\,$\mum, $\eta_0=50.5\,$mPa\,s, $V=4\,$m/s and the typical saturation deformation observed at very high concentration, $\beta \sim 3\times10^{-2}$ (see figure \reff{highphi}), the estimated dissipation of the dilatant motion, $P\Delta\phi/\phi\sim 10^4$\,Pa, is indeed found to be of the same order as the stagnation pressure $\rho V^2$. This scaling analyses, which neglects the details of the impact geometry and assumes uniform deformation and dilatation, is too simple to explain quantitatively all the experimental data. However, it explains qualitatively why the deformation remains controlled by $V$ and $\eta_0$ for $\phi\gtrsim\phic$. Furthermore, it predicts a crucial dependance of the deformation of the drop at large concentrations, namely, that the deformation should increase with increasing permeability of the solid phase, i.e., with increasing particle size. This dependance on $d$ is tested experimentally in figure \reff{d}{\it a}, which compares the spreadings measured for $d=585\,$\mum together with those obtained with $d=145\,$\mum for the same viscosity and impact velocity ($\eta_0 = 50.5\,$mPa and $V=4\,$m/s). Although the spreading is similar at low volume fractions ($\phi \lesssim 0.30$), it is found to be significantly larger for the larger particles at higher concentrations (both in terms of $\beta$ and of $\beta_c$). As illustrated more directly from the images in figure \reff{d}{\it b} for $\phi = 0.54$, for both $d=585$\,\mum and $d=145\,$\mum the two drops become more corrugated during the impact, which shows that in both cases the particle phase dilates. Yet, the drop with the larger particles deforms and spreads significantly more, as expected from a deformation limited by dilatancy.

\section{Conclusions}\label{sec:conclusion}

	We have considered the impact of a drop loaded with solid particles with a focus on the high volume fraction limit ($\phi\gtrsim\phic$), where a viscous effective approach predicts a vanishing deformation. A careful characterization of (i) the bulk effective viscosity of our suspensions and (ii) the low deformation limit of a viscous impact has provided us with a precise Newtonian effective reference. While the spreading of the suspension drop measured at moderate volume fractions is in agreement with the latter, a clear deviation is observed for larger values of $\phi$. In particular, a finite deformation is actually observed above $\phic$, in contradiction with the effective approach. This suggests that, although there is no splashing or particle ejection, the volume fraction of the suspension — in terms of the mean distance between the particles — is not fixed during the impact, which is consistent with the dilatant motion of the particle phase observed during the impact. The deformation around $\phic$ is found to depend on the particle size, not only above $\phic$ but also slightly below, with larger deformations observed for larger particles. This dependence cannot be attributed to a capillary effect related with the size of the menisci between the outermost particles, since the deformation is found to remain controlled by the viscosity of the suspending liquid for all values of $\phi$. This suggests that the observed dependence to the particle size is actually related with the dilatant motion itself, because dilatancy drives a Darcy flow that is all the more dissipative that the suspending liquid is viscous and the particles are small. This also means that a complete description of the impact of concentrated suspension drops will require to combine the unsteady surface flow facet of 
the problem with a two phases approach that can account for variations of the solid volume fraction and for the associated transient dissipation.
\vspace*{-3mm}

%

\bibliographystyle{jfm2}
\bibliography{Impact_JFM}

\end{document}